\documentstyle[preprint,aps,epsf]{revtex}
\begin{document}
\draft
\title{Equivalent frames in  Brans-Dicke theory}
\preprint{UTEXAS-HEP-00-5}
\author{Yungui Gong} 
\address{Physics Department, University of Texas at Austin, Austin, 
Texas 78712}
\maketitle
\begin{abstract}
We discuss the physical equivalence between
the Einstein and Jordan frames in Brans-Dicke theory.
The inequivalence of conformal transformed theories is clarified
with the help of an old equivalence theorem of Chrisholm's. 
\end{abstract}

Brans-Dicke theory is a natural generalization of Einstein theory.
In Brans-Dicke theory, the effective gravitational constant 
$G_{eff}=\phi^{-2}$, varies as the Brans-Dicke field $\phi$ evolves.
The evolution of the scalar field slows down the expansion rate
of the universe during inflation, and allows nucleation of bubbles
to end the inflationary era. There are two conformally related
frames, the Einstein frame and the Jordan frame, in Brans-Dicke 
theory. In the literature, people do not agree with each other
about the equivalence of the two frames. Some people think that the
two frames are physically equivalent and some people do not think
that they are physically equivalent. For example,
some people considered
the inflationary models in Einstein frame in order to solve equations
easily, but they analyzed their final results in Jordan frame 
because most people insist that the Jordan frame is the physical frame
to keep the equivalence principle. In fact, the equivalence principle
can be kept in Einstein frame if we use Einstein frame as the physical
frame. 
A dilaton field appears in Kaluza-Klein
reduction and string theory also. 
In string theory, we have the conformally related string frame and Einstein
frame. As we all know, the string frame is equivalent to the
Einstein frame. How could that the string frame is equivalent to the
Einstein frame and the Jordan frame is not equivalent to the Einstein frame?
In this letter, we will use the equivalence theorem to address this problem.
In general, we can ask if
the physics remains the same under an arbitrary change of field variables.
Let us first look at the equivalence theorem  \cite{cwz}\cite{ss}\cite{brst}.

{\bf Theorem}: Let the Lagrangian be known in terms of a set
of field variables $\phi$, $L[\phi]$. If one expresses the field
variables $\phi$ as nonlinear but local functions of another set of
field variables $\varphi$ (one may think of $\phi$ as a 
scalar field for simplicity), 
\begin{equation}
\label{equiv}
\phi=f[\varphi], \quad f[0]=0,\quad f'[\varphi]\neq 0,
\end{equation}
one can write down the Lagrangian as 
$L[\phi]=L[f(\varphi)]=L_t[\varphi]$.
The on-mass shell S matrices calculated with $L[\phi]$ and
$L_t[\varphi]$ are identical (in making the comparison it may be necessary
to introduce appropriate wave-function renormalizations).
In other words, both the fields $\phi$ and $\varphi$ can be 
used to describe the same physics.

The important point about the theorem is that 
we must keep the origin in the field
space unchanged
when we change the field variables so that
we can have the same free field Lagrangian
and we do not change the physical
quantities, like the mass of the field.
In gauge field theories, we give mass to the gauge fields
when we shift the Higgs fields by some constants.
In the above theorem, when we use the Lagrangian $L_t[\varphi]$,
we also need to introduce the Jacobian (the so called Lee-Yang term
\cite{leeyang}) due to the change of field variables. The contribution of
the Jacobian is necessary to cancel out the most divergent
term due to the derivative couplings. The Jacobian is also
necessary for an invariant measure in the field space \cite{ss}\cite{jhkm}.
However, it is not really too important because we set $\delta^4(0)=0$
when we use the dimensional regularization scheme. 

Now let us look at several examples. In string theory,
the Einstein frame is related with the string frame by
$g^E_{\mu\nu}=e^{-4\phi/(D-2)}g^{str}_{\mu\nu}$ and
the dilaton field $\phi$ is kept invariant. From the equivalence
theorem, we know that the string frame is physically equivalent
to the Einstein frame.

Suppose we have a free massless scalar
field theory,
\begin{equation}
\label{toya}
L=-{1\over 2}\partial_\mu\varphi\partial^\mu\varphi.
\end{equation}
Take $\varphi=\phi+\lambda\phi^3$, then the free massless
scalar field theory (\ref{toya}) becomes
\begin{equation}
\label{toyb}
L=-{1\over 2}\partial_\mu\phi\partial^\mu\phi-
3\lambda\phi^2\partial_\mu\phi\partial^\mu\phi-{9\over 2}
\lambda^2\phi^4\partial_\mu\phi\partial^\mu\phi.
\end{equation}
Now we get a massless scalar field theory with derivative self-interactions.
Since the coupling constant has mass dimension $-2$, the theory
in terms of $\phi$ is not power-counting renormalizable.
However, this nonrenormalizability is fictitious. We compute the
tree diagram for the four point and six point Green's function on
shell, the results are zero. The one loop contribution to
the $\phi-\phi$ scattering is also zero (the result is proportional
to $\delta^4(0)$ which is zero by dimensional regularization). 
These results are expected
because the theory described by (\ref{toyb}) is in fact a free theory
described by (\ref{toya}). 

If we take $\varphi=m_1 e^{\phi/m_1}$ instead, we get
\begin{eqnarray}
\label{toyc}
L&=&-{1\over 2}e^{2\phi/m_1}\partial_\mu\phi\partial^\mu\phi\nonumber\\
&=&-{1\over 2}\left[1+{2\phi\over m_1}+{2\phi^2\over m_1^2}
+{4\phi^3\over 3m_1^3}+\cdots\right]
\partial_\mu\phi\partial^\mu\phi.
\end{eqnarray}
Again we get a massless scalar field theory with 
derivative self-interactions.
It is not difficult to check that the tree and one-loop diagrams
give zero also. Although the exponential transformation does
not satisfy the condition of the equivalence theorem, the two
fields $\varphi$ and $\phi$ describe the same physics for this 
simple case. This may be understood from the fact that the free
field Lagrangians are the same. If we introduce self-interactions,
then it is obvious that the exponential transformation introduces
mass that changes the free field Lagrangian. Therefore, the 
exponential transformed theory is not physically equivalent
to the original theory in general.

However if we start from a free scalar field of
mass $m$,
\begin{equation}
\label{toy2}
L=-{1\over 2}\partial_\mu\varphi\partial^\mu\varphi
-{1\over 2}m^2\varphi^2.
\end{equation}
Take $\varphi=m_1 e^{\phi/m_1}$, then Eq. (\ref{toy2}) becomes
\begin{eqnarray}
\label{toy2a}
L&=&-{1\over 2}e^{2\phi/m_1}\partial_\mu\phi\partial^\mu\phi
-{1\over 2}m^2 m_1^2 e^{2\phi/m_1}\nonumber\\
&=&-{1\over 2}\partial_\mu\phi\partial^\mu\phi-m^2\phi^2-{1\over 2}m^2 m^2_1
-m^2 m_1\phi-{2m^2\over 3m_1}\phi^3-
{1\over m_1}\phi\partial_\mu\phi\partial^\mu\phi\nonumber\\
&&-{m^2\over 3m_1^2}\phi^4
-{1\over m^2_1}\phi^2\partial_\mu\phi\partial^\mu\phi+\cdots.
\end{eqnarray}
Now we have a scalar field of mass $\sqrt{2}m$ with self interaction.
The physical spectrum is changed. It is easy to verify that the tree
diagram for the four point Green's function is not zero on shell.

Now we are ready to discuss the Brans-Dicke theory.
The Jordan Brans-Dicke Lagrangian is given by
\begin{equation}
\label{bdlagr}
{\cal L}_{BD}=-{\sqrt{-\gamma}\over 16\pi}
\left[\phi{\tilde R}+\omega\,\gamma^{\mu\nu}
{\partial_\mu\phi\partial_\nu\phi\over \phi}\right]-{\cal L}_m(\psi,\,
\gamma_{\mu\nu}).
\end{equation}
If we expand the metric $\gamma_{\mu\nu}$ around
the flat metric $\eta_{\mu\nu}$, ${\tilde h}_{\mu\nu}=\gamma_{\mu\nu}-
\eta_{\mu\nu}$ does not represent the spin-2 massless graviton. Instead,
it is $\rho_{\mu\nu}={\tilde h}_{\mu\nu}+a\,\sigma\eta_{\mu\nu}$ that
represents the spin-2 massless graviton. 
The Lagrangian (\ref{bdlagr}) is conformal invariant under
the conformal transformations,
$$g_{\mu\nu}=\Omega^2\gamma_{\mu\nu},\quad \Omega=\phi^\alpha,~~
(\alpha\neq {1\over 2}),\quad \sigma=\phi^{1-2\alpha}.$$
For the case $\alpha=1/2$, we make the following transformations
\begin{mathletters}
\begin{equation}
\label{conformala}
g_{\mu\nu}=e^{a\sigma}\gamma_{\mu\nu},
\end{equation}
\begin{equation}
\label{conformalb}
\phi={8\pi\over \kappa^2}e^{a\sigma},
\end{equation}
\end{mathletters}
where $\kappa^2=8\pi G,\  a=\beta\kappa,\ \beta^2={\displaystyle
2\over \displaystyle 2\omega+3}$. 
After the conformal transformations (\ref{conformala})
and (\ref{conformalb}),
we get the Einstein  Brans-Dicke Lagrangian 
\begin{equation}
\label{lagr}
{\cal L}= \sqrt{-g} \left[-\frac{1}{2\kappa^2}{\it R}
-\frac{1}{2}g^{\mu\nu}\partial_\mu \sigma \partial_\nu \sigma\right]
-{\cal L}_{m}(\psi, e^{-a\sigma}g_{\mu\nu}).
\end{equation}
Note that the Brans-Dicke Lagrangian is not
invariant under the transformations (\ref{conformala}) and
(\ref{conformalb}).
In this frame, $h_{\mu\nu}=g_{\mu\nu}-\eta_{\mu\nu}$
represents the spin-2 massless graviton. That's the reason why
we think the Einstein frame is the physical frame.
The kinetic term of the $\phi$ field in Eq. (\ref{bdlagr}) can
be rewritten as the familiar form $\gamma^{\mu\nu}\partial_\mu\phi
\partial_\nu\phi$ if we let $\phi=\varphi^2$. As discussed in
the examples, the transformations (\ref{conformala}) 
and (\ref{conformalb}) give two physically inequivalent theories
if the scalar field $\phi$ has potential term. For the case
discussed here (Eq. (\ref{bdlagr}) and Eq. (\ref{lagr})), the equivalence
theorem tells us that the Einstein frame is equivalent to the Jordan
frame because the free field Lagrangians are the same. 
In other words, if we have the same in states 
$|\gamma_{in}\rangle$ and $|g_{in}\rangle$ and out states
$\langle\gamma_{out}|$ and $\langle g_{out}|$, then the cross
section calculated in the Jordan frame using the Lagrangian 
(\ref{bdlagr}) is the same as that calculated in the Einstein frame
using the Lagrangian (\ref{lagr}). However, the meaning of the
equivalence between the Jordan frame and the Einstein frame in the
literature is not what we stated here. For example, cosmological
models in the Jordan frame are different from those in the
Einstein frame. The reason is very simple. If we use the same
physical parameters, like the Hubble constant at the present,
in both the Jordan frame and the Einstein frame, then the in and
out states are not the same in the two frames. In \cite{bdej},
there are some explicit examples to show the difference
between the two frames. 

The equivalence theorem can tell us
if two theories written in terms of different field variables are equivalent. 
If two theories written in terms of different field variables are not
physically equivalent, the theorem can not tell us which theory should be
thought as the physical one. If we consider the interactions between
gravity and matter fields, then the physical condition, namely the
Einstein equivalence principle,
rules out the conformal transformations (\ref{conformala}). 
If we impose a minimal coupling between 
the gravity and matter
in the original theory, then we lose the minimal coupling 
after the conformal transformations.
In other words, we can have covariant conservation 
law $\nabla^\nu T_{\mu\nu}$ in the
original theory, but we do not have the covariant 
conservation law in the transformed theory.

\acknowledgments

The author would like to thank Professor Yuval Ne'eman and 
Bryce DeWitt for fruitful discussions.


\begin{references}
\bibitem{cwz}  J.S.R. Chisholm, Nucl. Phys. {\bf 26}, 469 (1961);
S. Kamefuchi, L. O'Raifeartaigh and A. Salam, {\it ibid.} {\bf 28}, 529 (1961);
S. Coleman, J. Wess and B. Zumino, Phy. Rev. {\bf 177}, 2239 (1969);
R.E. Kallosh and I.V. Tyutin, Sov. J. Nucl. Phys. {\bf 17}, 98 (1973);
Y.-M. P. Lam, Phys. Rev. D {\bf 7}, 2943 (1973);
M.C. Bergere and Y.-M. P. Lam, {\it ibid.}, {\bf 13}, 3247 (1976);
M. Bando, T Kugo and K. Yamawaki, Phys. Rep. {\bf 164}, 217 (1988);
J. Donoghue, E. Golowich and B. Holstein, {\it Dynamics of the Standard Model} (Cambridge
University Press, Cambridge, 1992), Page 101.
\bibitem{ss} A. Salam and J. Strathdee, Phys. Rev. D {\bf 2}, 2869 (1970).
\bibitem{brst} A.A. Slavnov, Phys. Lett. {\bf 258B}, 391 (1991);
F. Bastianelli, Nucl. Phys. {\bf B361}, 555 (1991);
J. Alfaro and P.H. Damgaard, Annals Phys. {\bf 220}, 188 (1992);
A. Blasi, N. Maggiore, S.P. Sorella and L.C.Q. Vilar, 
Phys. Rev. D {\bf 59}, 121701, (1999);
I.V. Tyutin, hep-th/0001050 (2000).
\bibitem{leeyang} T.D. Lee and C.N. Yang, Phys. Rev. {\bf 128}, 885 (1962).
\bibitem{jhkm} J. Honerkamp and K. Meetz, Phys. Rev. D {\bf 3}, 1996 (1971);
I. S. Gerstein, R. Jackiw, B. W. Lee and S. Weinberg, {\it ibid.}
2486 (1971); S. Weinberg, {\it The Quantum Theory of Fields} (Cambridge
University Press, Cambridge, 1995), Page 389-393.
\bibitem{bdej} Y. Gong and Y.Z. Zhang, Europhys. Lett. {\bf 31}, 7 (1995); 
Y. Gong, gr-qc/9809015, unpublished (1998); 
Phys. Rev. D {\bf 59}, 083507 (1999).
\end{references}
\end{document}